\documentclass[11pt]{article} 
\usepackage[utf8]{inputenc}

\usepackage{amsfonts}
\usepackage{amsmath}
\usepackage{amssymb}
\usepackage{euscript,array}
\usepackage[dvipsnames]{xcolor}
\usepackage{cancel}
\usepackage{amsthm}

\usepackage{psfrag}
\usepackage{graphicx}
\usepackage{color}
\usepackage{stmaryrd}
\usepackage{cancel}

\usepackage{tikz}
\usepackage{verbatim}

\usetikzlibrary{arrows,shapes,backgrounds} 
\usepackage{tikzpagenodes} 
\usepackage{fancyhdr}

\usepackage[process=auto]{pstool}

\usepackage{caption}
\usepackage{color}
    \definecolor{darkgreen}{rgb}{0,0.5,0}
    \definecolor{darkblue}{rgb}{0,0,0.6}
    \definecolor{purple}{rgb}{0.4,.2,0.7}
\usepackage[margin = 2.5cm]{geometry}
\usepackage{graphicx}
\usepackage[hyperfootnotes = false, colorlinks = true, linkcolor = darkblue, citecolor = purple]{hyperref}
\usepackage{subcaption}
 
\newcommand{\be}{\begin{equation}}
\newcommand{\ee}{\end{equation}}
\newcommand{\bea}{\begin{eqnarray}}
\newcommand{\eea}{\end{eqnarray}}

\begin{document}

\thispagestyle{empty}
\begin{center}
    ~\vspace{5mm}
    
    {\LARGE \bf {Remarks on the large-$N$ \boldmath ${\mathbb C}P^{N-1}$ model\\}} 
    
    \vspace{0.5in}
    
    {\bf   Antonino Flachi}

    \vspace{0.5in}

    Department of Physics, \&\\ Research and Education Center for Natural Sciences,\\ Keio University, Hiyoshi, Yokohama, Japan \vskip1em
     
    \vspace{0.5in}
    
     {\tt   flachi@phys-h.keio.ac.jp}

\end{center}

\vspace{0.5in}

\begin{abstract}
In this paper we consider the ${\mathbb C}P^{N-1}$ model confined to an interval of finite size at finite temperature and chemical potential. We obtain, in the large-$N$ approximation, a mixed-gradient expansion of the one-loop effective action of the order parameter associated with the effective mass of the quantum fluctuations. This expansion gives an expression for the thermodynamic potential density as a functional of the order parameter, generalizing previous calculations to arbitrarily large order and to the case of finite chemical potential and allows one to discuss some generic features of the ground state of the model.
The technique used here relies on analytic regularization and provides an efficient scheme to extract the coefficients of the expansion. Once a solution for the ground state is known these coefficients can be used to deduce some generic properties of the ground state as a function of external conditions. 
We also show that there can be no transition to a massless phase for any value of the external conditions considered and clarify a seemingly important point regarding the regularization of the effective action connected to the appearance of logarithmic divergences and to the Mermin-Wagner-Hoenberg-Coleman (MWHC) theorem. 
\end{abstract}

\vspace{1in}

\pagebreak

\setcounter{tocdepth}{3}


\section{Introduction}
\label{sec:intro}

The ${\mathbb C}P^{N-1}$ model is $1+1$ dimensional ($d=1$) field theory, consisting of $N$ complex scalar fields $n_i$ ($i=1,2,\cdots, N$) with an action of the form 
\bea
\EuScript S = \int dx dt \left| D_{\mu} n_i \right|^2,
\label{eqt1}
\eea 
with $D_\mu = \partial_\mu -i A_\mu$, with the U(1) gauge field $A_\mu$ lacks at classical level a kinetic term (in principle, a kinetic term may reappear at one-loop). The fields $n_i$ are forced to obey a constraint,
\bea
\left| n_i \right|^{2} = r.
\label{eqt2}
\eea 
Original work on the model dates back at least to References~\cite{Polyakov:1975rr,Polyakov:1975yp,Bardeen:1976zh,Brezin:1976qa,DAdda:1978vbw,Witten:1978qu} (see Refs.\cite{Zinn-Justin:2002,Shifman:2012} for textbook introductions), but recent years have seen a resurgence of interest in the properties of its ground state when the model is confined to an interval of finite size $\ell$ ($x \in \left[0,\ell\right]$) and fluctuations subjected to boundary conditions or other external forcing, as for example temperature variations. Refs.~\cite{Hong:1994,Hong2:1994} were the first (to the best of our knowledge) to look into questions related to this confined setup, and since then a renewed interest and a very active debate have resurfaced (see, for example, Refs.~\cite{Shifman:2007rc,Gorsky:2013rpa,Monin:2015xwa,Milekhin:2016fai,Milekhin2:2016fai,Bolognesi:2016zjp,Pikalov:2017lrb,Pikalov:2020lrb,Flachi:2017xat,Nitta3:2018yen,Pavshinkin:2017kwz,Nitta:2018yen,Nitta2:2018yen,Betti:2017zcm,Bolognesi:2018njt,Bolognesi:2019rwq,Chernodub:2019nct,Gorsky:2018lnd,Flachi:2019jus,Ishikawa:2019tnw,Ishikawa:2019oga}) leading to various, and sometimes contradictory, claims being made. While the above references refer to large-$N$ calculation, the ${\mathbb C}P^{N-1}$ model has also been the subject of extensive lattice simulations \cite{Berg:1981er,Campostrini:1992ar,Alles:2000sc,Beard,Bruckmann:2015sua,Flynn:2015uma,Kawauchi:2016xng,Abe:2018loi,Bonanno:2018xtd,Fujimori:2019skd,Berni:2019bch,Pelissetto:2019iic} (see in particular Ref.~\cite{Fujimori:2019skd} for a lattice study of the ${\mathbb C}P^{N-1}$ model on $\mathbb{S}^1 \times \mathbb{S}^1$). Issues being currently debated have to do with whether the model can develop a massless ground state for small enough interval size, how the properties of the ground state depend on the external conditions (i.e., size, boundary conditions and temperature), and how everything fits under the umbrella of the large-$N$ approximation. 

Our goal here is to reexamine the story and extend the analysis to the case of finite density. We are motivated by two main objectives. The first one is related to the possibility that inhomogeneous phases, even if energetically disfavored at zero density, may become  favored above a critical density. This is known to happen for the Gross-Neveu, Nambu-Jona Lasinio and quark-meson models (see, for example, Refs.~\cite{Nakano:2004cd,Nickel:2009ke,Nickel2:2009ke,Boehmer:2007ea,Abuki:2011pf}) and it is quite reasonable to expect a similar situation occurring for the ${\mathbb C}P^{N-1}$ model. This may be interesting since it could lead to new features in the geography of the phase diagram of the model (that is the appearance of crossovers into regions characterized by inhomogeneous phases), even for the case of periodic boundary conditions\footnote{Although technically non-trivial, it is obvious to expect, away from periodic boundary conditions, the ground state to become spatially modulated.}. The se\-cond reason is to inspect whether a transition from a massive to a massless phase may or may not occur and whether there is any clear mechanism to exclude the existence of a massless phase (both possibilities have been entertained in the literature with differing conclusions; see Ref.~\cite{Bolognesi:2019rwq} and references given there). Extending the calculation to finite density gives us an excuse to reconsider this debated matter. 

The paper is structured as follows. In Sec. \ref{sec2}, we introduce the main setup and notation, and illustrate the calculation of the effective action at finite temperature, density, and size using zeta-regularization. This calculation is essentially a repetition of that of Ref.~\cite{Flachi:2019jus} with two major differences: the first being the inclusion of a chemical potential, and the second being a different regularization that allows to capture the infrared behavior of the model and leads to the appearance of a logarithmic contribution. This is an issue of some importance, since it is this term that eventually prevents a massless ground state to be realized and locks the system into a massive phase. Because of this, in Sec. \ref{sec3}, we will show how the presence of logarithmic contributions can be understood on rather general grounds using zeta-function regularization. This will be done by exploiting the analytic structure of the zeta-function associated with the problem by means of its \textit{Mittag-Leffler} representation in general dimensionality. In Sec. \ref{sec4}, we will discuss the implications of the calculation for the ground state.
Some formulas involving polylogarithmic functions used in the computation are given in Appendix \ref{appendix}.

 \section{One-loop effective action at finite chemical potential}
\label{sec2}

{
In order to examine the effect of one-loop quantum effects, we shall proceed by incorporating the constraint (\ref{eqt2}) in the tree-level action (\ref{eqt1}) by means of a Lagrange multiplier $M^2$, leading to the following expression
\bea
\EuScript S = \int dx dt \left(\left| D_{\mu} n_i \right|^2 + M^2 \left(\left| n_i \right|^{2} - r\right)\right).
\label{eqtaux1}
\eea 
Variation of the action with respect to $M^2$ enforces the constraint. The quantity $M^2$ plays the role of an \textit{effective} mass and it is initially assumed to be in principle a spatially varying function; minimization of the effective action will then determine whether a constant or inhomogeneous configuration will be realized.
}

{
In the present work we set $A_\mu=0$. While this is consistent with the choice of periodic boundary conditions, in general, the U(1) gauge symmetry of the ${\mathbb C}P^{N-1}$ model may in general be broken by different boundary conditions. Therefore fixing the gauge field \textit{a priori} (rather than through minimization of the (effective) action) may be inconsistent with some choice of boundary conditions or viewed as a restricting assumption. In the present case, our focus is on periodic boundary conditions that do not break such gauge invariance and allow us to set the gauge field to zero. 
The same choice has been made in previous works (See, for example, Refs.~\cite{Bolognesi:2016zjp,Bolognesi:2018njt,Bolognesi:2019rwq,Flachi:2019jus}).
}

Here, we follow Ref.~\cite{Flachi:2019jus} and perform a coordinate transformation, $x \to \tilde x = x/\ell$ and $\tau \to \tilde \tau = \tau/\ell$ ($\tau$ is the Wick-rotated Euclidean time and $\beta=1/T$ in the expression above represents the inverse temperature), in order to rescale the interval to one of unit length. These rescaled coordinates are dimensionless and we use the symbol $\tilde \nabla \left(= \ell \nabla\right)$ to indicate differentiation with respect to the rescaled coordinate $\tilde x$. In the following we choose the background-field configuration along the $k=1$ direction, i.e., $n_k = \sigma \times \delta_{1k}$ with $k=1,2,\cdots, N$ with $\delta_{ik}$ being the Kronecker delta. This yields for the one-loop effective action at large-$N$ the following expression
\bea
\EuScript  S^{E}_{\mbox{\tiny{eff}}} &=& 
\int_0^{\beta/\ell} d \tilde\tau \int_0^1 d \tilde x 
\left\{ 
\left({\tilde \nabla} \sigma\right)^2  + \ell^2 M^2 \left( \left| \sigma \right|^{2} - r
\right) {- \ell^2\mu^2 \sigma^2}
\right\} + \delta \Gamma,
\eea
where the quantity $\delta \Gamma$ is the one-loop determinant 
\bea
\delta \Gamma
= {(N-1)\over 2} \sum_\pm \mbox{Tr} \log \left(-\tilde\Delta -{\partial^2\over \partial\tau^2}+ \ell^2 M^2 {- \ell^2\mu^2\pm 2 \ell^2\mu {\partial \over \partial \tau}}\right).
\label{eqt3.1}
\eea
The above expression for the one-loop effective action at large-$N$ at finite temperature and chemical potential is readily obtained after path-integration over the fields $n_k$ and $n^*_k$ and, for $\mu=0$ coincides with those of Refs.~\cite{Bolognesi:2016zjp,Bolognesi:2018njt,Bolognesi:2019rwq,Flachi:2019jus}.
As explained at the beginning of this section, the constraint (\ref{eqt2}) has been incorporated by means of a Lagrange multiplier $M^2$ (as $\delta S/\delta M^2=0$) that operates as an effective mass. 
{While we use throughout the paper the terminology ``massive" and ``massless" phase or ground state, these correspond to the ``Coulomb/confinement" phase ($M^2 \ne 0$, $\sigma =0$) and ``Higgs (or
deconfinement) phase" ($M^2=0$, $\sigma \ne 0$), respectively.}
The quantities $M^2$ and $\sigma$ are assumed to be {time-independent}, but otherwise general functions of space. The sum over the functional determinant goes over both $\pm$ signs 
\cite{Haber:1981ts,Toms:1992dq,Toms2:1992dq}.

{Here, we follow Ref.~\cite{Bruckmann:2014sla} and introduce a chemical potential $\mu$ associated with the first component of the complex $n_i$, as this is analogue to a chemical potential associated to a $U(1)$ symmetry of a free complex scalar field. This is the simplest possible choice, and despite the fact that it is not the most general configuration, it is sufficient to understand whether at finite chemical potential the ground state of the model (defined as the background field $\sigma$ and the Lagrange multiplier $M^2$ that extremize the effective action and have lowest free energy) may acquire a spatial dependence. More complex configurations (chemical potentials coupled to other or all conserved charges) can be seen as a combination of several elementary configurations as discussed in Ref.~\cite{Bruckmann:2014sla}, but these come in at a price of more involved calculations. Also, here we are focusing on the regime of $\mu$ not large; addressing what happens at large values of the chemical potential is certainly worth of attention and will be considered elsewhere.}
Taking the limits of $\mu \to 0$ and $M\to 0$ in the previous formulas recuperate, respectively, known formal expressions (See, for example, Refs.~\cite{Flachi:2019jus,Haber:1981ts,Toms:1992dq,Toms2:1992dq}). 

Using zeta-regularization, we can express the effective action in terms of the zeta-function (see Refs.~\cite{elizalde94,Avramidi,ParkerToms} for textbook derivations)
\bea
\zeta(s)= \sum_{k=0}^\infty\sum_{n=-\infty}^\infty \left(p^{(s)}_{k} + \left(2\pi n\ell/\beta {\pm i \ell\mu}\right)^2\right)^{-(D-d)}
\label{eqt4}
\eea
at $D-d =s$, as 
\bea
\delta \Gamma = -\zeta_\pm'(0).  
\label{eqt5}
\eea 
The (dimensionless) eigenvalues $p^{(s)}_{k}$ are defined by 
$$
\left(\tilde\Delta_s +\ell^2 M^2\right) f_k = \left(p^{(s)}_{k}\right)^2 f_k
$$ 
and encode the dependence on $ M$, $\ell$ and on the boundary conditions. The operator $\tilde\Delta_s$ is the regularized version of $\tilde\Delta = \lim_{s\to 0} \tilde\Delta_s$ (similarly to what is done in dimensional regularization, here we analytically continue the dimensionality, $d \to D=d+s$, and let $s\to 0$ at the end). 

The computation of the derivative of the zeta function can be performed in the usual way by utilizing the Mellin transform, 
\bea
a^{-s} \Gamma(s) = \int_0^\infty t^{s-1} e^{-a t} dt,
\label{eqt6}
\eea 
to re-express the zeta function (\ref{eqt4}) in terms of the (integrated) heat-kernel $\EuScript K_s(t)$ (defined below) associated to the operator $\left(\tilde\Delta_s +\ell^2  M^2\right)$. Simple calculations give
\bea
\zeta_\pm(s)&=& 
{1\over \Gamma(s)} 
\int_0^\infty {dt\over t^{1-s}}
\EuScript K_s(t)
 \sum_{n=-\infty}^\infty  e^{-(\Omega_n^\pm)^2 t},
\label{eqt7}
\eea 
where we have defined 
\bea
\Omega_n^\pm = 2\pi n\ell/\beta {\pm i \ell\mu}
\label{eqt8}
\eea 
and
\bea
\EuScript K_s(t) = \sum_{k} e^{-t \left(p^{(s)}_{k}\right)^2},
\label{eqt9}
\eea
where $\EuScript K_s(t) $ represents the heat-kernel in $D=d+s$ dimensions (In the present case, $d=1$ and the regularization parameter $s$ is let to zero at the end of the calculations). 

The expression of the zeta function can be rearranged by using the following identity: 
\bea
 \sum_{n=-\infty}^\infty  e^{-(\Omega_n^\pm)^2 t}
= {\beta/\ell \over \sqrt{4 \pi t}} \vartheta_3\left(\pm{i\beta \mu\over 2},e^{-{\beta^2 \over 4 \ell^2 t}}\right),
\label{eqt10}
\eea 
where $\vartheta$ is a Jacobi theta function \cite{Abramowitz}. Using Eq.~(\ref{eqt10}) in Eq.~(\ref{eqt7}) we get
\bea
\zeta_\pm(s)&=& 
{1\over \Gamma(s)} {\beta/\ell \over \sqrt{4 \pi}} 
\int_0^\infty {dt\over t^{3/2-s}}
\EuScript K_s(t) \times 
\vartheta_3\left(\pm{i\beta \mu\over 2},e^{-{\beta^2 \over 4 \ell^2 t}}\right).~~~~~~
\label{eqt11}
\eea
To evaluate the derivative of the zeta function and the effective action, we express the integrated heat kernel in terms of the heat-kernel density, ${\EuScript K}_s(t)=\int dx\; {\EuScript K}_s(x, t)$ and use the following small-$t$ expansion
\bea
{\EuScript K}_s(x, t) = {
1 \over \left({4 \pi t}\right)^{d+s\over 2}} \sum_{k=0}^\infty \tilde{\alpha}^{(s)}_{k} t^{k}.
\label{eqt12}
\eea
The first four coefficients reduce in the limit $s\to 0$ to (see, for example, Ref.~\cite{Bastianelli:2008}): 
\bea
\tilde{\alpha}^{(0)}_0 &=&1,\nonumber\\
\tilde{\alpha}^{(0)}_1 &=& - \ell^2 M^2,\nonumber\\
\tilde{\alpha}^{(0)}_2 &=& {1\over 2}\ell^4 M^4 - {1\over 6}\tilde \Delta \left(\ell^2 M^2 \right)
, \nonumber\\
\tilde{\alpha}^{(0)}_3 &=& 
- {1\over 6} \ell^6 M^6 + {1\over 12} \left(\tilde \nabla \left(\ell^2 M^2\right)\right)^2 + {1\over 6} \ell^2 M^2 \tilde \Delta \left(\ell^2 M^2 \right)
- {1\over 60} \tilde \Delta^2 \left(\ell^2 M^2 \right).
\nonumber
\label{eqt13}
\eea
{The above heat-kernel expansion is a derivative expansion and it is valid when higher order derivatives become less important, that is when the function $M^2$ is not a rapidly varying function of the spatial coordinate. We stress that this is an assumption here, based on the physical intuition that rapidly varying functions are expected to have a higher energy. However, in the limit of small size, one may expect that this approximation breaks down. The present approach does not allow to determine what is the scale below which this approximation breaks down. A direct computation of the effective action under the different assumption that $M^2$ is indeed a rapidly varying function is, in principle, possible, but we will not consider it here.}

{
Proceeding in this way (details are given below) yields the bulk part of the effective action from which the gap equation (i.e., the equation for the function $M^2$) can be obtained. This part of the effective action is \textit{independent} of the boundary conditions (i.e., it is valid for \textit{any} choice of boundary conditions). As for the boundary conditions that one needs to impose on the function $M^2$, these follow from the boundary conditions imposed on the fields $n_k$. These induce an additional (boundary) contribution to the effective action, which determines how $M^2$ behaves at the boundary.\\ 
In general, the boundary part of the effective action can be obtained following an identical procedure, once the boundary contribution to the heat-kernel coefficients is included. 
For periodic boundary conditions, that is our focus here, such a ``boundary" part vanishes (in the case of periodic boundary conditions, there is no boundary). For other choices of boundary conditions (i.e., Dirichlet, Neumann, Robin, coupled), the boundary part is nonvanishing and will result in a nontrivial condition for $M^2$ at the boundary.
In the present case, we are interested in periodic case and therefore simply ignore the boundary contribution. Should one be interested in other boundary conditions (say, leading to Dirichlet or Neumann or other boundary conditions for $M^2$), one can take the bulk equation (obtained from the effective action obtained here) and solve under the added requirement that any solution must have the appropriate boundary behavior. Away from periodic boundary conditions, such solutions will be inhomogeneous. In the case of periodic boundary conditions, both solutions (constant or inhomogeneous) are possible and the usual expectation is that the constant solution is the lower energy one. For the model at hand, this point has been debated recently \cite{Gorsky:2018lnd,Bolognesi:2019rwq}. 
}

Returning to the computation of the effective action, the next step to carry out the integration over $t$ conveniently, we express the theta function using the following series representation:
\bea
\vartheta_3\left(x,y\right) = 1 + 2 \sum_{n=1}^\infty \cos (2 n x) y^{n^2},
\label{eqt14}
\eea
which allows us to write the zeta-function as follows
\bea
\zeta_\pm(s)&=& 
{1\over \Gamma(s)} {\beta/\ell \over {(4 \pi)^{(d+1+s)/2}}} 
\sum_{k=0}^\infty \tilde{\alpha}^{(s)}_{k} 
\int_{\Lambda^{-2}}^\infty {dt\over t^{(d+3-2k-s)/2}} \left(1 + 2 \sum_{n=1}^\infty \cosh \left(\beta \mu n\right) e^{-{\beta^2n^2 \over 4 \ell^2 t}}
\right),~~~~~~
\label{eqt15}
\eea
with the limit $\Lambda \to \infty$ understood. It is a good point to remark that in our dimensionless coordinates, the parameter $\Lambda$ is also dimensionless. Dimension-full quantities can be reintroduced by transforming back to the original coordinate system, as we shall do later. Assuming $\Re s$ to be sufficiently negative and proceeding by analytical continuation, the integrals over $t$ can be performed exactly giving
\bea
\zeta_\pm(s)&=& 
{\beta/\ell \over {(4 \pi)^{(d+1+s)/2}}\Gamma(s)}
\sum_{k=0}^{k_\star} \tilde{\alpha}^{(s)}_{k}
\left( -{\Lambda^{-s -2k +d+1} \over 
s/2  +k - (1+d)/2} \right.\nonumber\\
&&\left.+2 \Gamma\left((1+d)/2-k-s/2\right)
\left({\beta \over 2\ell}\right)^{s+2k-d-1}
\sum_{n=1}^\infty
\cosh \left(\pm \beta \mu n\right) n^{s+2k-d-1}
\right),\nonumber
\eea
where the sum over $k$ extends to finite but arbitrarily large $k=k_\star$. Defining $z=\beta \mu$ and 
\bea
\varpi_\pm\left(a\right) =
\sum_{n=1}^\infty
\cosh \left(\pm z n\right) n^{-1+a},
\label{eqt16}
\eea
and noticing that any term with $k \geq 2$ is regular in the limit $s \to 0$ allows us to write in the limit $d\to 1$
\bea
\lim_{s\to 0}{d \zeta_\pm \over d s} 
&=& {\beta/\ell \over 4 \pi}
\left\{
\Omega^{\pm}_0
+
\left[{1\over 2} \tilde\alpha_1^{(0)} 
\left(
-\gamma_E+\log \pi 
+2 \log \left({\beta\over\ell}\right)
+2 \log \Lambda^2 - 4 \varpi_\pm'(1)
\right)- \lim_{s\to 0}{d \alpha_1^{(s)} \over ds}\right]
\right.\nonumber\\
&&\left.+
\sum_{k=2}^{k_\star}
{(-1)^k\over \Gamma(k) 2^{2k-3}} {\beta^{2k-2}\over \ell^{2k-2}} \varpi_\pm'(2k-1) \tilde \alpha_k^{(0)} 
\right\}.
\nonumber
\eea
In arriving at the above expression, we have used the following relations (these and the other relations involving the functions $\varpi_\pm$ used here are derived in Appendix)
\bea
\varpi_\pm(3) &=& \varpi_\pm(5)=0, \label{eqtpollo1}\\
\varpi_\pm(1)&=&-1/2,\label{eqtpollo2}
\eea
and we have defined the quantity (for $d\to1$)
\bea
\Omega^{\pm}_0 &=& 
\lim_{s\to 0}{d \over d s}
{\tilde{\alpha}^{(s)}_{0}\over{(4 \pi)^{s/2}}\Gamma(s)} 
\left[-{\Lambda^{-s+2} \over s/2  - 1}
+\Gamma\left(1-{s\over 2}\right)\left({\beta \over 2\ell}\right)^{-2+s}
\varpi_\pm \left(s-1\right)
\right],
\nonumber
\eea
that is a divergent vacuum energy contribution, independent of $M^2$ in the limit $s\to 0$. We can now write
\bea
\zeta'_+(0) + \zeta'_-(0)
&=&{\beta/\ell \over 4 \pi}
\left\{
\delta \Omega_0
+
\left[\tilde\alpha_1^{(0)} 
\left(
-\gamma_E+\log \pi 
+2 \log \Big({\beta\over\ell}\right)\right.
+2 \log \Lambda^2 - 2 \varpi'(1)
\Big)- 2\lim_{s\to 0}{d \tilde\alpha_1^{(s)} \over ds}\right]\nonumber\\
&&\left.+
\sum_{k=2}^{k_{\star}}
{(-1)^k\over \Gamma(k) 2^{2k-3}} {\beta^{2k-2}\over \ell^{2k-2}} \varpi'(2k-1) \tilde \alpha_k^{(0)} 
\right\},
\label{eqt18.1}
\eea
where we have defined
\bea
\delta \Omega_0 &=&   \Omega^+_0+\Omega^-_0, \nonumber\\
\varpi(z) &=& \varpi_+(z)+\varpi_-(z).
\eea
The above results can be combined to arrive at the following expression for the one-loop effective action:
\bea
\EuScript  S^{E}_{\mbox{\tiny{eff}}} &=& 
{\beta}\int_0^\ell d x 
\left\{ 
\left(\nabla\sigma\right)^2  + M^2 \left( \left| \sigma \right|^{2} - r_\star
\right)-\mu^2\sigma^2 
- {(N-1) \over {4 \pi}} 
\left[\delta \Omega_0
-
\left(
\log \left({\beta^2\over\ell^2}\right)
- 2 \varpi'(1)
\right)M^2
\right.
\right.
\nonumber\\
&&
\left.
\left.
-M^2 \log\left(\ell^2 M^2\right)
+ {\beta^2\over 4} {}  \varpi'(3) M^4
+ {\beta^4 \over 16} {} \varpi'(5) \left({1\over 6}  M^6 + {1\over 12} \left( \nabla \left(M^2\right)\right)^2 \right)
+\cdots
\right]
\right\},
\label{eqt19}
\eea
after appropriately reabsorbing terms proportional to $M^2$ with constant coefficients and divergences into a renormalized coupling $r_\star$ and after eliminating total derivatives. For $\mu \to 0$, formulas
\bea
\omega_\pm(a) = \zeta_R(1-a)\; ~~~~ \mbox{and}\;~~~~
\omega(a) = 2 \zeta_R(1-a),
\eea
allow us to straightforwardly recover the result of Ref.~\cite{Flachi:2019jus}, with the exception of the logarithmic contribution $M^2\log\left(\ell^2M^2\right)$ present here. This term arises from the contribution $\lim_{s\to 0}{d \tilde\alpha_1^{(s)} / ds}$ in formula (\ref{eqt18.1}) and originates from the regularization of the differential operator in (\ref{eqt3.1}) analytically continued from $d$ to $D=d+s$. Then, the heat-kernel coefficients associated with the regularized operator scale with the mass as in $D=d+s$ dimensions. In the present case, $d=1$, the only nontrivial contribution comes from $\tilde a_{(1+d+s)/2}$. This term scales as $\tilde a_{(1+d+s)/2} = \left(\ell M\right)^{1+d+s}$ that leads, in the effective action in $d=1$, to the logarithmic term, $M^2 \log\left(\ell^2 M^2\right)$, in (\ref{eqt19}). All higher ($k\geq 2$) order contributions are regular in the limit $s \to 0$, while the $k=0$ contribution is divergent but $M^2$ independent, thus only resulting in a constant shift in the energy. 
The logarithmic contribution is quite important in $1+1$ dimensions since it is a manifestation of the Mermin-Wagner-Hoenberg-Coleman theorem \cite{Mermin,Hohenberg,Coleman}(or, reversing the logic, in the present setup the restrictions of the theorem follow from this term that encodes an infrared diverging behavior in the $M\to 0$ limit). This is readily seen once the constraint $\delta \EuScript  S^{E}_{\mbox{\tiny{eff}}} /\delta M^2=0$ is implemented: the logarithmic correction yields a singularity, as $\log M^2$, impeding any solution with $M^2 =0$ to be realized and thus excluding any massless phase from the spectrum. Once the Lagrange multiplier is integrated out in the path integral, such a zero mode must then be excluded. This conclusion seems to be perfectly in tune with that of Ref.~\cite{Bolognesi:2019rwq} and the additional term is the missing ingredient that brings to an agreement the results of Refs.~\cite{Bolognesi:2019rwq,Flachi:2019jus}.

 \section{Logarithmic contributions and the Mittag-Leffler representation}
\label{sec3}

The presence of the logarithmic contribution discussed in the preceding section can be understood on rather general grounds and quite easily in zeta-function regularization. 

Here, we limit our consideration to a second order differential operator of Laplace type $\EuScript D = g^{\mu\nu} \nabla_\mu \nabla_\nu + E$ in $D=d+\epsilon$ spatial dimensions, where $\nabla_\mu$ is a suitable covariant derivative and $E$ is an endomorphism. The covariant derivative may include gauge potentials or connection due to external fields or spacetime curvature, and our consideration below are valid in general. The case considered in the previous section refers to the one-dimensional Laplacian operator with $E=M^2$. The one-loop effective action $\Gamma \left(\EuScript D\right)$ can be formally written as \cite{ParkerToms}
\bea
\Gamma = \sum_\lambda \log \left( \hat{\mu}^{-2}\lambda\right), 
\label{zetafct}
\eea
where the summation over the eigenvalues $\lambda$ of $\EuScript D$ is understood as a regularized sum. In (\ref{zetafct}) we have assumed that the eigenvalues have $\left[mass^2\right]$ dimension and introduced an arbitrary (renormalization) constant 
$\hat\mu$ to keep the argument of the logarithm dimensionless. Introducing the following zeta-function
\bea
\zeta\left(\epsilon|\EuScript D \right) = \sum_\lambda \left(\hat\mu^{-2} \lambda\right)^{-\epsilon},
\label{z1}
\eea 
allows one to write the above one-loop determinant as follows
\bea
\Gamma \sim \lim_{\epsilon\to 0}\zeta'\left(\epsilon|\EuScript D \right),
\label{z2}
\eea 
where the limit is understood in the sense of analytical continuation. Now, it is possible to prove that if the operator $\EuScript D$ is positive definite, then the zeta function is amenable of an expansion of the form \cite{Kirsten:2001,Gilkey:1984}
\bea
\zeta\left(\epsilon|\EuScript D \right) = {1\over \Gamma(\epsilon)} \left[\sum_{p=0}^\infty {\alpha_p (\EuScript D) \over p-(D+1)/2} + \EuScript F(\epsilon)\right]
\label{z3}
\eea
known as \textit{Mittag-Leffler} expansion [the assumption of a strictly positive operator can be relaxed to a non-negative operator with modified coefficients in the numerator of (\ref{z3})]. In the above expression, the quantities $\alpha_k (\EuScript D)$ are the heat-kernel coefficients associated to the operator $\EuScript D$ and $\EuScript F(\epsilon)$ is an entire function. {Let us focus here on the case of $D$ odd (in the case of $D$ even, and in the absence of boundaries $\alpha_{(D+1)/2}=0$. This does not clash with the MWHC theorem, since its restrictions do not apply in dimensions higher than 3). Thus in $D+1$ (even) spacetime dimensions}, $\alpha_{(D+1)/2} (\EuScript D)$ is the heat-kernel coefficient responsible for the divergences and it scales as
\bea
\alpha_{(D+1)/2} (\EuScript D) \sim E^{(D+1)/2} + \cdots,
\label{z4}
\eea 
where in flat space and in absence of external gauge potentials the dots denote mixed-derivative terms that vanish in the limit $E$ constant. The term (\ref{z4}) above is sufficient to deal with the present situation of $d=1$. In higher dimensionality in the presence of curvature of gauge potentials additional terms (not vanishing in the limit of $E \to$ constant) need to be accounted for \cite{Kirsten:2001,Gilkey:1984}, but the argument given here does not change. Then, using (\ref{z1}), (\ref{z3}), (\ref{z4}), the presence of the logarithm becomes apparent:
\bea
\Gamma &\supset& E^{(1+d)/2} \log \left(E/\mu^2\right).
\eea
In the preceding section we have been concerned with the case of $d=1$, $E=M^2$ and $\hat\mu=\ell^{-1}$, leading precisely to the $M^2\log \ell^2M^2$ term appearing in (\ref{eqt19}). These results have interesting physical implications for the Casimir effect and will be presented elsewhere \cite{Flachi:2020pvn}.

\section{Discussions}
\label{sec4}

With the results of the preceding sections in hands, we can examine some of the features of the ground state of the model. 

As we have already mentioned, the presence of the logarithm $M^2\log \ell^2M^2$ yields a $\log M^2$ divergence once the constraint, $\delta \EuScript  S^{E}_{\mbox{\tiny{eff}}} /\delta M^2=0$, is implemented, impeding the realization of a massless phase. This result is independent of the external conditions, that is a massless ($M^2=0$) phase \emph{cannot} be realized by increasing the density, the temperature or decreasing (or \emph{increasing}) the size of the interval. This is nothing but the manifestation of the Mermin-Wagner-Hoenberg-Coleman theorem \cite{Mermin,Hohenberg,Coleman} that becomes evident in the analytic regularization we have used here. Importantly, this also shows that there is no clash between the restrictions of the theorem and the large-$N$ approximation. This is reminiscent of Refs.~\cite{Witten:1978qu}.


{In order to understand (to the present level of approximation) whether a spatially modulated $M^2$ is energetically favored can be understood directly from the form of the effective action, similarly to Ref.~\cite{Nakano:2004cd,Nickel:2009ke,Nickel2:2009ke,Boehmer:2007ea,Abuki:2011pf}. However, two things should be kept in mind. As discussed in Ref.~\cite{Bolognesi:2019rwq}, when the effective action is extremized with respect to a Lagrange multiplier (and assuming that no kinetic term is generated at one- or higher-loop order), then one should require that the extrema of the effective action is a maximum (and not a minimum). The second and more important point is that 
despite the fact that both constant and inhomogeneous solutions are in principle allowed and consistent with periodic boundary conditions (the EOM admit both solutions), the argument of Ref.~\cite{Bolognesi:2019rwq} yields a proof of uniqueness of the ground state to be spatially homogeneous. In this case, the spatial homogeneity of the ground state indicates that the derivative terms are not only sub-leading, but in fact vanishing for the ground state solution of  Ref.~\cite{Bolognesi:2019rwq}.}

The question that remains is whether for more general boundary conditions, for which case the bulk part of the effective action takes the same form as derived here, there exist multiple inhomogeneous solutions. In this case, our expansion together with the argument of Ref.~\cite{Bolognesi:2019rwq} gives a criteria to select which of the solutions maximize or minimize the effective action and therefore can be accepted as ground state.

\begin{figure}[t]
\centering
\begin{tabular}{cc}
\includegraphics[scale=0.7]{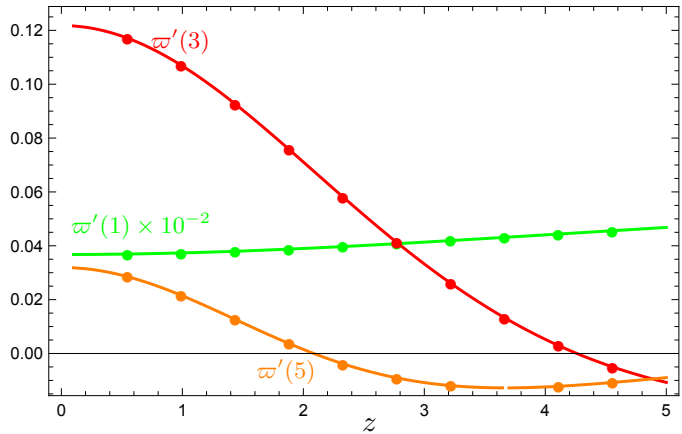}
\put(-120,-5){{$z$}}
\put(-230,70){\footnotesize{\textcolor{green}{$\varpi'(1)\times 10^{-2}$}}}
\put(-160,15){\footnotesize{\textcolor{orange}{$\varpi'(5)$}}}
\put(-200,140){\footnotesize{\textcolor{red}{$\varpi'(3)$}}}
\end{tabular}
\caption{\label{figura} Profiles of the coefficients $\varpi'(1)$, $\varpi'(3)$ and $\varpi'(5)$ as a function of $z=\beta \times \mu$ calculated using the representation (\ref{dervarpi}) (continuous curves) and numerically starting from the definition of polylogarithmic function points (dots). The agreement has been verified up to accuracy of $10^{-7}$.}
\end{figure}

Practically, to inspect whether a spatially modulated solution may become energetically disfavored, as external conditions are varied, we need to compute the dependence of the coefficients 
$\varpi'(1)$, $\varpi'(3)$ and $\varpi'(5)$ on the temperature and on the chemical potential (in the present case, the relevant coefficients do not depend on the size, $\ell$).
This can be easily done either by using formula (\ref{dervarpi}) derived in appendix, or by \textit{brute force} numerical computation, starting from the definition of polylogarithmic functions \cite{Abramowitz}. (Using this second approach will result in an imaginary part for the function $\varpi$  due to lack of choice in performing the analytical continuation in our numerical scheme. The imaginary part is then discarded and the real part compared with the result obtained from  formula (\ref{dervarpi}) that gives a real value.) We have carried out the computation in both ways (numerics were carried out with an accuracy of $10^{-7}$) and compared the results that perfectly agreed. Results are shown in Fig.~\ref{figura}. We should remark that the expansion has been carried out to order 6 in the heat-kernel expansion. This implies that the parameters of the expansion stay small (that is the combination of temperature and chemical potential accompanying higher order terms are small enough to be ignored), otherwise additional terms in the expansion have computed. This is in fact straightforward to do in our scheme, requiring only the evaluation of higher order $\omega'(z)$ coefficients. 

In the present case, since no massless phase can be realized, there is no phase transition. Then, the sign of the coefficient of the $M^2$ term simply dictates the gradient of the potential for small $M^2$. The coefficient of $M^4$, that in absence of the logarithm would determine the order of the transition (and a change from second order for $\varpi'(3)>0$ to first order for $\varpi'(3)<0$), here simply controls the concavity of the potential. The coefficient $\varpi'(5)$ is instead more meaningful since it is the first term in the expansion (\ref{eqt19}) multiplying a derivative contribution
and thus signaling when spatially modulated solutions (when they exist) are energetically favored or disfavored.  
For $z \gtrsim z_{crit} \approx 2.05$, $\varpi'(5)$ turns negative, indicating a decreases in the free energy, keeping homogeneous solutions favored. For periodic boundary conditions, this is the only possibility, as it follows from the uniqueness argument of Ref.~\cite{Bolognesi:2019rwq}. 

Another point worth noticing is the independence of the coefficients of all powers of $M^2$ in the expansion from the size of the interval (with the exception of the logarithms). While this can be explicitly observed from formula (\ref{eqt19}) for the $M^4$ and $M^6$ coefficients, a proof that extends to all coefficients is worked out very easily from formula (\ref{eqt18.1}) and from the scaling of the coefficients $\tilde \alpha^{(0)}_k$. This is a reminder of the large-$N$ volume independence for the $\mathbb{C}P^{N-1}$ model (see \cite{Sulejmanpasic:2016llc}).

\section{Conclusions}
 
In this paper we have examined a number of issues, recently debated (see Refs.~\cite{Shifman:2007rc,Gorsky:2013rpa,Monin:2015xwa,Milekhin:2016fai,Milekhin2:2016fai,Bolognesi:2016zjp,Flachi:2017xat,Pikalov:2017lrb,Nitta3:2018yen,Pavshinkin:2017kwz,Nitta:2018yen,Nitta2:2018yen,Betti:2017zcm,Bolognesi:2018njt,Bolognesi:2019rwq,Chernodub:2019nct,Gorsky:2018lnd,Flachi:2019jus,Ishikawa:2019tnw,Ishikawa:2019oga}), on the features of the ground state of the $\mathbb{C}P^{N-1}$ model at finite temperature, (small) density and size. We have worked out an expansion \emph{\`a la} Ginzburg-Landau for the effective action as a functional of the Lagrange multiplier $M^2$, that enforces the constraint on the fundamental fields of the model and ope\-rates as an effective mass. Assuming $M^2$ to be in principle spatially varying, the coefficients of the expansion easily allow one to determine whether there is any phase transition as temperature, density and size vary. Using analytical continuation based on zeta-function regularization, we have been able to show that a logarithmic term of the form $M^2 \log \left(\ell^2 M^2\right)$ occurs in the one-loop effective action. This term yields a divergent contribution once the constraint is implemented, preventing the realization of a massless phase in complete agreement with the Mermin-Wagner-Hoenberg-Coleman theorem. To summarize, our calculations indicate:
\begin{itemize}
\item the absence of a massless phase for any value of the external conditions (therefore no phase transition towards a massless phase);
\item at vanishing density, derivative terms increase the energy of the ground state, therefore inhomogeneous phases are energetically disfavored;
\item our expansion along with the uniqueness of the ground state (as shown in Ref.~\cite{Bolognesi:2019rwq}) implies that the ground state is always spatially homogeneous at any density, temperature, and size (when $\beta \mu \leq 2 \pi$).
\end{itemize}

In all of the above we have taken periodic boundary conditions. Beyond periodic boundary conditions, the ground state naturally develops spatial inhomogeneities. While the bulk part of the effective action we have derived here is still valid, the addition of boundary terms must be included and the boundary part of the action can be easily worked out for the present setup following a procedure similar to Ref.~\cite{Flachi:2019jus}. Whether multiple solutions are possible and transitions between inhomogeneous grounds states may occur remains to be seen. 

In conclusion, we should remark that the scheme presented here is limited by the validity of the derivative expansion (that is in essence an expansion in powers of $\beta/ \ell$) and by the series representation of the function $\varpi$ (that is valid for $\beta \mu \leq 2 \pi$). It would certainly be desirable to improve the results of this paper in order to be able to extend the present expansion to the case of large chemical potential, a problem that requires finding the correct analytical continuation of the $\varpi'(z)$ coefficients beyond the case studied here. Beyond the case of periodic boundary conditions it may also be interesting to look at rapidly varying solutions, that can, in principle, be done by re-summing certain classes of derivative terms in the heat-kernel expansion. Another interesting point concerns the interplay between the restrictions resulting from Mermin-Wagner-Hoenberg-Coleman theorem and the Casimir force, particularly in dimensions greater than $2$. We hope to report on these in forthcoming work \cite{Flachi:2020pvn}.

\textit{Acknowledgements.} Thanks are due Enzo Vitagliano for encouraging comments and questions that ultimately led to this work; to T. Fujimori, G. Marmorini, and K. Ohashi for generously sharing their insight on nonlinear sigma models; to my collaborators on \cite{Flachi:2019jus}, Muneto Nitta, Satoshi Takada, and Ryosuke Yoshii, for many discussions on related topics and to Guglielmo Fucci for discussions on polylogarithms. Finally, I should like to thank Kenji Fukushima for clarifications regarding the sign problem and the large-$N$ expansion. The support of the Japanese Ministry of Education, Culture, Sports and Science (MEXT-supported Program for the Strategic Research Foundation at Private Universities "Topological Science" Grant No.\ S1511006) and of the Japanese Society for the Promotion of Science (Grants-in-Aid for Scientific Research KAKENHI Grant n. 18K03626) are gratefully acknowledged.

\appendix
\section{Series representation of the function $\varpi(s)$.}
\label{appendix}

In order to compute the coefficients $\varpi'(1)$, $\varpi'(3)$ and $\varpi'(5)$, we shall start from the following expression
\bea
f\left(a; x\right) =
2 \sum_{n=1}^\infty
\cosh \left(x n\right) n^{-1+a}
= 
\mbox{Li}_{1-a}\left(e^{-x}\right)+\mbox{Li}_{1-a}\left(e^{x}\right),
\label{app1}
\eea
where $0 \leq \left| x \right|< 2\pi$. Using the expression above and tabulated values of polylogarithmic functions \cite{Abramowitz}, it is easy to verify that 
\bea
\omega_\pm(1) = -1/2.
\label{app2}
\eea
Using the following identity
\bea
\mbox{Li}_{1-n}\left(e^{-z}\right)+ (-1)^n \mbox{Li}_{1-n}\left(e^{z}\right) = 0,
\label{app3}
\eea
with $n\in \mathbb{N}$, it follows that
\bea
{\omega_\pm(3) = \omega_\pm(5) = 0.}
\label{app4}
\eea
formulas (\ref{app2}) and (\ref{app4}) are those used in (\ref{eqtpollo1}) and (\ref{eqtpollo2}).

To compute the coefficients $\varpi'(p)$, we shall adopt the following series representation of the polylogarithmic function:
\bea
{\mbox{Li}_{1-a}\left(e^{x}\right) = \Gamma(a) \left(-x\right)^{-a} + \sum_{k=0}^\infty {1\over k!}{\zeta_R}\left(1-a-k\right) x^k},
\label{appzapp}
\eea 
valid for $a\in \mathbb{C}/\mathbb{N}$ and $\left|x\right| \in (0,2 \pi)$ \cite{Robinson:1951}. In the domain $1-a \leq 0$ with $a\notin \mathbb{N}$, the representation above is an analytic function and the series converge absolutely for all $\left|x\right| \leq 2 \pi$. For $a\in \mathbb{N}$, it is possible to extend the domain by analytical continuation.  The properties of the above series representations have been discussed in various references (see, for example, \cite{Fucci:2011mf} and the list of references given there). Using the above relation (\ref{appzapp}) and defining
\bea 
x_\pm = \pm \left|x\right|
\eea
we can easily arrive at 
\bea
\varpi(a)
&=& \varpi_+(a) + \varpi_-(a) \nonumber\\
&=&
2 \Gamma(a) \left| x \right|^{-a} \left(1 + \cos \left(\pi a\right)\right)
+ 4 \sum_{k=0}^\infty {1\over (2k)!}{\zeta_R}\left(1-a-2k\right)\left|x \right|^{2k},
\eea
where we have have analytically continued $\varpi_+(a)$ from the top and $\varpi_-(a)$ from the bottom. From the above expression is easy to obtain for the coefficients $\varpi'(a)$ the following formula:
\bea
\varpi'(a)
= - 4 \sum_{k=0}^\infty {1\over (2k)!}{\zeta_R'}\left(1-a-2k\right)\left|x \right|^{2k}.
\label{dervarpi}
\eea
The above representation is regular and can be compared against a \textit{brute force} numerical computation carried out using the \textit{definition} of the polylogarithmic function. Finally, notice that for $z=0$ (corresponding to $\mu=0$) we have
\bea
{\omega'(a) = -4{\zeta'_R}\left(1-a\right)}.
\nonumber
\eea

\end{document}